\documentclass[times,twocolumn,final]{elsarticle}

\usepackage{cag}
\usepackage{framed,multirow}

\usepackage{amssymb}
\usepackage{latexsym}
\usepackage{amsmath}

\usepackage{url}
\usepackage{xcolor}
\definecolor{newcolor}{rgb}{.8,.349,.1}

\usepackage{hyperref}

\usepackage[switch,pagewise]{lineno} 

\newcommand{\eg}{\emph{e.g.}}
\newcommand{\ie}{\emph{i.e.}}
\definecolor{newcolor}{rgb}{.8,.349,.1}
\newcommand{\revise}[1]{{\color{black}{#1}}}

\journal{Computers \& Graphics}

\begin{document}

\verso{Preprint submitted for review}

\begin{frontmatter}

\title{UrbanVR: An immersive analytics system for context-aware urban design}%

\author[1]{Chi Zhang}
    
\author[2]{Wei Zeng}
\emailauthor{wei.zeng@siat.ac.cn}{Wei Zeng}

\author[1]{Ligang Liu}

\address[1]{University of Science and Technology of China, Hefei, 230026, China}
\address[2]{Shenzhen Institute of Advanced Technology, Chinese Academy of Sciences, Shenzhen, 518055, China}

\received{\today}

\begin{abstract}
Urban design is a highly visual discipline that requires visualization for informed decision making.
However, traditional urban design tools are mostly limited to representations on 2D displays that lack intuitive awareness.
The popularity of head-mounted displays (HMDs) promotes a promising alternative with consumer-grade 3D displays.
We introduce UrbanVR, an immersive analytics system with effective visualization and interaction techniques, to enable architects to assess designs in a virtual reality (VR) environment.
Specifically, UrbanVR incorporates 1) a customized parallel coordinates plot (PCP) design to facilitate quantitative assessment of high-dimensional design metrics, 2) a series of egocentric interactions, including gesture interactions and handle-bar metaphors, to facilitate user interactions, and 3) a viewpoint optimization algorithm to help users explore both the PCP for quantitative analysis, and objects of interest for context awareness.
Effectiveness and feasibility of the system are validated through quantitative user studies and qualitative expert feedbacks.
\end{abstract}


\end{frontmatter}


\section{Introduction} \label{sec:intro}
Visualization plays a key role in supporting informed discussions among stakeholders in an urban design process:
For designers, the design process comprises visual representations in practically all stages and for most aspects $-$ from ideation and specification to analysis and communication~\cite{burhard_2007_visualization};
For decision makers, visualizations help to understand implications of the design~\cite{delaney_2000_visualization}.
Specifically, the process of designing a city’s development site is rather tedious: designers need to come up with several design options, then analyze them against key performance indicators, and finally select one.
The process requires effective analysis and visualization tools.

Urban environments are composed of buildings and various surroundings, which can be naturally represented as 3D models. 
As such, many 3D visual analytics systems have been developed for a variety of urban design applications, including cityscape and visibility analysis~\cite{ferreira_2015_urbane, ortner_2016_vis-a-ware}, flood management~\cite{waser_2014_many, cornel_2015_visualization}, vitality improvement~\cite{zeng_2018_vitalvizor}, and recycling~\cite{richthofen_2017_urban}.
Yet, most of them are developed for desktop displays, which lack \emph{spatial presence} that is the sense of ``being there" in the world depicted by the virtual environment~\cite{slater_1999_measuring}.
The 3D nature of urban environments calls for immersive analytics tools that can provide vivid presence of the urban environment beyond the desktop.

The advance of affordable, consumer head-mounted displays (HMDs) such as the HTC Vive, has revived the fields of VR and immersive analytics.
Recently, there is a growing interesting in utilizing VR technology for urban planning, which can improve design efficiency and facilitate public engagement~\cite{Liu_2020_three, schrom2020interactive}.
Nevertheless, developing such an immersive analytics system is challenging:
First, the system needs to present the urban environment (3D spatial information) together with mostly quantitative analysis data (abstract information) that are typically high-dimensional.
It remains challenging to display both spatial and abstract information in the virtual environment~\cite{chen_2017_exploring}.
Second, conventional interactions for desktop displays, such as mouses and keyboard, are infeasible for VR.
New ways to interact with the immersive analytics systems are required~\cite{ens_2021_grand}.

We present UrbanVR, an immersive analytics system that integrates advanced analytics and visualization techniques to support the decision-making process in site development in a VR environment.
UrbanVR caters to various analytical tasks that are feasible for analysis and visualization in VR, identified from semi-structured interviews with a collaborating architect (Sect.~\ref{sec:overview}).
We then focus on visualization and interaction design for supporting site development in a virtual environment.
Specifically, we design a parallel coordinates plot (PCP) that can be spatially situated with 3D physical objects, to facilitate quantitative assessment of high-dimensional analysis metrics (Sect.~\ref{ssec:vis_design}).
Next, we integrate a series of egocentric interactions, including gesture interactions, and handle bar metaphors, to facilitate user interactions in VR (Sect.~\ref{ssec:inter_design}).
We further develop a viewpoint optimization algorithm to mitigate occlusions and help users explore spatial and abstract information (Sect.~\ref{ssec:viewport}).
A quantitative user study demonstrates effectiveness of these interactions, and qualitative feedbacks from domain experts confirm the applicability of UrbanVR in supporting shading and visibility analysis (Sect.~\ref{sec:eva}).

The primary contributions of our work include:
\begin{itemize}
\item UrbanVR is an immersive analytics system with a 3D visualization of an urban site for context-aware exploration, and a customized parallel coordinate plot (PCP) for quantitative analysis in VR.

\item UrbanVR integrates viewport optimization and various egocentric interactions such as gesture interactions and handle-bar metaphors, to facilitate user exploration in VR.

\item A quantitative user study together with qualitative expert interviews demonstrate the effectiveness and applicability of UrbanVR in supporting urban design in virtual reality.
\end{itemize}
\section{Related Work} \label{sec:related}

This section summarizes previous studies closely related to our work in the following categories.

\vspace{1.5mm}
\noindent
\emph{Visual Analytics for Urban Data}.
Nowadays, big urban data are being ubiquitously available, which has promoted the development of evidence-based urban design.
Meanwhile, visualization has been recognized as an effective means for communication and analysis.
As such, many visual analytics systems have been developed to support various urban design applications, like transportation planning~\cite{zeng_2014_visualizing} and environment assessment~\cite{shen_2017_streetvizor}.
Most of these systems utilize 2D maps for visualizing big urban data~\cite{zheng_2016_visual}, which omits 3D buildings and surroundings in an urban site.
Recent years have witnessed some visual analytics systems for urban data in 3D.
For instance, VitalVizor~\cite{zeng_2018_vitalvizor} arranges 3D visualization of physical entities and 2D representation of quantitative metrics side-by-side for urban vitality analysis.
Closely related to this work, Urbane~\cite{ferreira_2015_urbane}, Vis-A-Ware~\cite{ortner_2016_vis-a-ware}, and Shadow Profiler~\cite{8283638} provide 3D visualizations for comparing effects of new buildings on landmark visibility, sky exposure, and shading.
Our work also caters to landmark visibility and shadow analysis.
But instead of presenting 3D spatial information on 2D display, we render 3D visualizations in a VR environment, to enable more vivid experience.

\vspace{1.5mm}
\noindent
\emph{Immersive Analytics}.
VR HMDs provide an alternative to 2D displays for data visualization in an immersive environment.
Immersive visualization is naturally appropriate for spatial data, \eg, scientific data~\cite{virtual_1996_steve}.
Advances of technical features of VR HMDs including higher resolution and lower latency tracking, promote a wide adoption of consumer-grade HMDs for data visualization and analytics in VR, \ie, \emph{immersive analytics}~\cite{dwyer_2018_immersive}.

Nevertheless, there are many challenges when developing immersive analytics systems.
The challenges can be grouped into four topics: spatially situated data visualization, collaborative analytics, interacting with immersive analytics systems, and user scenarios~\cite{ens_2021_grand}.
Recently, many immersive visualization techniques for displaying abstract information have been developed and evaluated, such as DXR toolkit~\cite{sicat_2019_dxr}, glyph visualization~\cite{chen_2019_marvist}, flow maps~\cite{yang_2019_od}, heatmaps~\cite{perhac_2017_urban, kraus_2020_assessing}, and tilt map~\cite{yang_2020_tilt}.
These techniques provide expressive visualizations for building immersive analytics.
For example, DXR toolkit~\cite{sicat_2019_dxr} provides a library of pre-defined visualizations such as scatterplots, bar charts, and flow visualizations.
Yet, these visualizations are infeasible for displaying the high-dimensional shadow and visibility analysis metrics.
As such, we design a customized PCP that can be overlaid on top of a building, to address the challenge of spatially situated data visualization.


\vspace{1.5mm}
\noindent
\emph{Interaction Design}.
Interactivity is one of the key elements of a vivid experience in VR~\cite{sherman_2003_interacting}.
Advanced interaction techniques are required for simultaneous exploration of 3D urban context and abstract analysis metrics.
Egocentric interaction that can embed users in VR is the most common approach for immersive visualizations~\cite{poupyrev_1998_egocentric}.
We design interactions as follows to facilitate interaction with immersive analytics systems.
\begin{itemize}

\item
\emph{Direction Manipulation}.
VR HMDs block users' vision of the real environment, thus traditional interactions (\eg, mouse and keyboard) for 2D displays are not applicable anymore.
On the other hand, mid-air gestural interaction can mimic the physical actions we make in the real world, which has been studied as a promising approach to 3D manipulation~\cite{besancon_2021_star}.
For example, Huang et al.~\cite{huang_2017_gesture} developed a gesture system for abstract graph visualization in VR.
Moreover, virtual hand metaphors have been studied for enhancing 3D manipulation, for both multi-touch displays~\cite{sun_2013_multi-touch} and mid-air~\cite{song_2012_handle}.
In line with these studies, this work presents a systematical categorization and development of gesture interactions required for immersive urban development, which will allow users to manipulate 3D urban context and interact with abstract analysis metrics.

\item
\emph{Viewpoint Motion Control}.
Occlusion is considered as a weakness of 3D visualization~\cite{teyseyre_2009_overview}.
In the development stage, we observed that users often spend much time on changing their viewpoint when manipulating 3D physical objects, for mitigating the occlusion problem.
To facilitate the navigation process, Ragan et al.~\cite{ragan_2017_amplified} designed a head rotation amplification technique that maps the user's physical head rotation to a scaled virtual rotation.
Alternatively, viewpoint optimization techniques have been proven effective by extensive studies in visualization~\cite{1532833} and computer graphics~\cite{https://doi.org/10.1111/j.1467-8659.2004.00781.x}. A principle metric for these methods is view complexity~\cite{plemenos:hal-00080226}, and the metrics are expanded with visual saliency, view stability, and view goodness~\cite{https://doi.org/10.1111/j.1467-8659.2008.01181.x}. 
As such, we develop an automatic approach for viewpoint optimization to reduce navigation time.
\end{itemize}
\section{Overview}\label{sec:overview}
In this section, we discuss domain problems derived from a collaborating architect (Sect.~\ref{ssec:requirements}), followed by a summary of distilled analytical tasks (Sect.~\ref{ssec:task}). 

\subsection{Domain Requirements}
\label{ssec:requirements}
In a six-month collaboration period, we worked closely with an architect having over 10 years of experience in urban design. The architect is often asked to propose building development schemes for a development site, which mainly involves considerations of the following entities.
\begin{itemize}
	\item \emph{Static environment:} Physical entities surrounding the development site. The entities cannot be modified.
	\item \emph{Building candidates:} The buildings that are proposed by architects to be placed on the development site. Their sizes and orientations can be modified.
	\item \emph{Landmark:} A physical entity in the development site that is easily seen and recognized.
\end{itemize}

Quantitative evaluation of the building candidates need to be carried out, in order to meet the evidence-based design principle. Many evaluation criteria have been proposed, such as visibility, accessibility, openness, etc. After evaluation, the building candidates and corresponding evaluation criteria will be demonstrated to stakeholders.
\begin{itemize}
\item \emph{Visualization Requirements}. The expert showed us existing tools for site development, and pointed out the lack of intuitive visualizations. He expressed a strong desire of utilizing emerging VR HMDs, as stakeholders are usually impressed with new technologies. The visualization should include both 3D geometries of the development site, and abstract information of analysis results.

\item \emph{Analytics Requirements}. Together with the expert, we identified two key criteria that should be carefully evaluated in site development: visibility and shadow. Visibility analysis is applicable to a landmark, while shadow analysis for the static environment. Both criteria are preferably visualized in an immersive environment.

\item \emph{Interaction Requirements}. Architects working with city planning and development often manually manipulate physical models with hands. As such, the architects prefer to interact with virtual objects using gestural interactions, instead of HMD coupled controllers that they are not familiar with. The collaborating architect also suggested to support system exploration without body movement to reduce the risk of falling, stumbling over cables, or inducing motion sickness.
\end{itemize}

\subsection{Task Abstraction}
\label{ssec:task}
The overall goal of this work is to develop an immersive analytics system for accessing site development. Based on the domain requirements, we follow the nested model for visualization design and validation~\cite{munzner_2009_nested}, and compile a list of analytical tasks in terms of data/operation abstraction design, and encoding/interaction technique design:
\begin{itemize}
	\item \emph{T.1: Quantitative Analysis.} The system should provide quantitative analysis of the identified criteria, \ie, visibility (T.1.1) and shadow (T.1.2). Efficient computation algorithms should be developed such that the analyses can be performed interactively.
	\item \emph{T.2: Multi-Perspective Visualization.} The system should present multi-perspective information in an immersive environment. First, a 3D view of the urban area should be provided to enable context-aware exploration (T.2.1). Second, an analytics view should be presented to visualize quantitative results (T.2.2).
	\item \emph{T.3: Effective Interaction.} The system should be coupled with a robust gesture interaction system (T.3.1). Actions on virtual objects should be directly visible to the users (T.3.2). Lastly, the system should provide optimal viewpoints to reduce body movement (T.3.3).
\end{itemize}
\section{UrbanVR System} \label{sec:sys}
The UrbanVR system consists of three components: data analysis (Sect.~\ref{ssect:data_ana}), visualization design (Sect.~\ref{ssec:vis_design}), and interaction design (Sect.~\ref{ssec:inter_design}).

\begin{figure}[!t]
\centering
\includegraphics[width=0.485\textwidth]{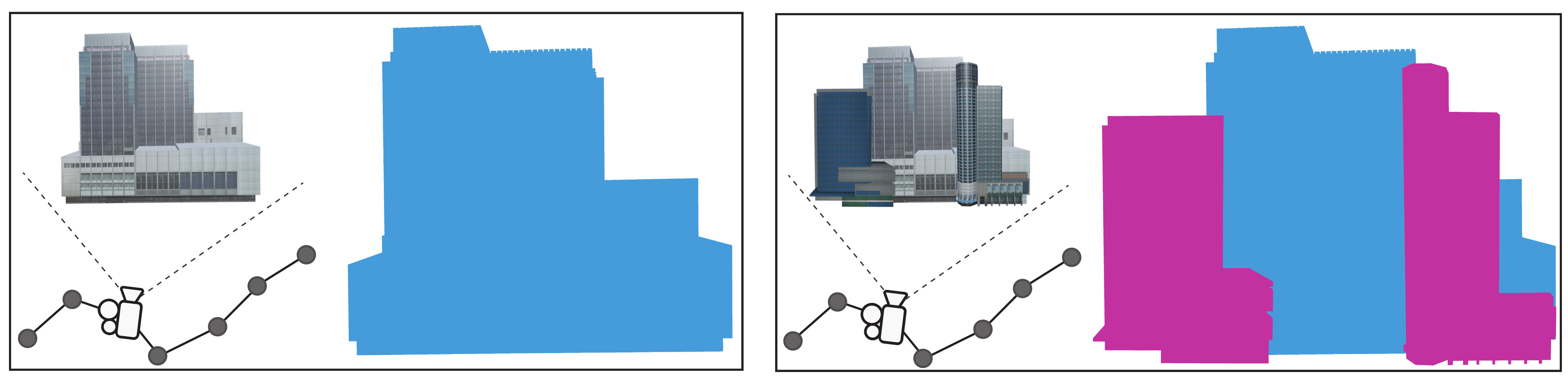}
\caption{Illustration of visibility analysis along a path of viewpoints using image analysis.}
\label{fig:anaa}
\end{figure}

\subsection{Data Analysis}
\label{ssect:data_ana}
The analysis component supports the quantitative analysis task (T.1). Here, both shadow and visibility measurements are calculated using GPU-accelerated image processing methods. The analyses include the following steps:
\begin{itemize}
	\item \emph{Viewpoints Generation.} Visibility and shadow are measured for a landmark and static environment, respectively. Both measurements are performed along a path of viewpoints in accordance to real-world scenarios, e.g., a city tourist tour and sun movement over one day. Specifically, viewpoints for visibility analysis are manually specified by the collaborating architect, which simulates a popular route in the urban area. Viewpoints for shadow analysis are automatically generated using a solar position algorithm\footnote{https://midcdmz.nrel.gov/spa/}.
	\item \emph{Image Rendering.} For each viewpoint, two high-resolution images are rendered. First, we create a frame buffer object (FBO) denoted as FBO-1, and render only the target in blue color to FBO-1 (Fig.~\ref{fig:anaa}(left)). Then, we create FBO-2 and render the entire urban scene to it, with the target rendered in blue color and other models in red color (Fig.~\ref{fig:anaa}(right)). For both images, the camera is directed towards the center of the target.

	\item \emph{Pixel Counting.} We can then extract the amount of blue pixels, denoted as $PC^b_1$ and $PC^b_2$ in FBO-1 and FBO-2, respectively. Then we compute visibility as $PC^b_2 / PC^b_1$. For shadow analysis, a building candidate produces shadows on the static environment. Thus, the shadow brought in by a building candidate is computed as $1-PC^b_2/PC^b_1$. To accelerate the computation, we develop a parallel computing method that divides each image into $10\times10$ grids and counts blue pixels in each grid using GPU.

	An alternative approach to the above described pixel counting is to directly clip other models when rendering the target, which could be even faster. Yet the image processing method can be generalized to other urban design scenarios, such as isovist analysis that measures the volume of space visible from a given point in space.
\end{itemize}

At each viewpoint, we precompute these measurements for each building in $\pi/3$ increments and at three scales. In this way, we generate a total of 18 (6 orientations $\times$ 3 scales) design variations for each building candidate. The shadows from 8:00 to 18:00 for each design are measured every 30 minutes. Our system loads the precomputed results for visualization. At runtime, our system also allows users to manipulate a building candidate, such as to scale the building up, or change its orientation. After the interaction, the system will recompute the measurements in the background. With the GPU-accelerated image processing method and this background computing approach, our system achieves interactive frame rates at runtime.
Note that visibility analysis is an approximation of stereographic projection of a landmark onto VR glasses. This approximation is acceptable since human eyes distance is much smaller than the viewpoint sampling step, which is about 20 meters in our work.

\begin{figure}[!t]
\centering
\includegraphics[width=0.485\textwidth]{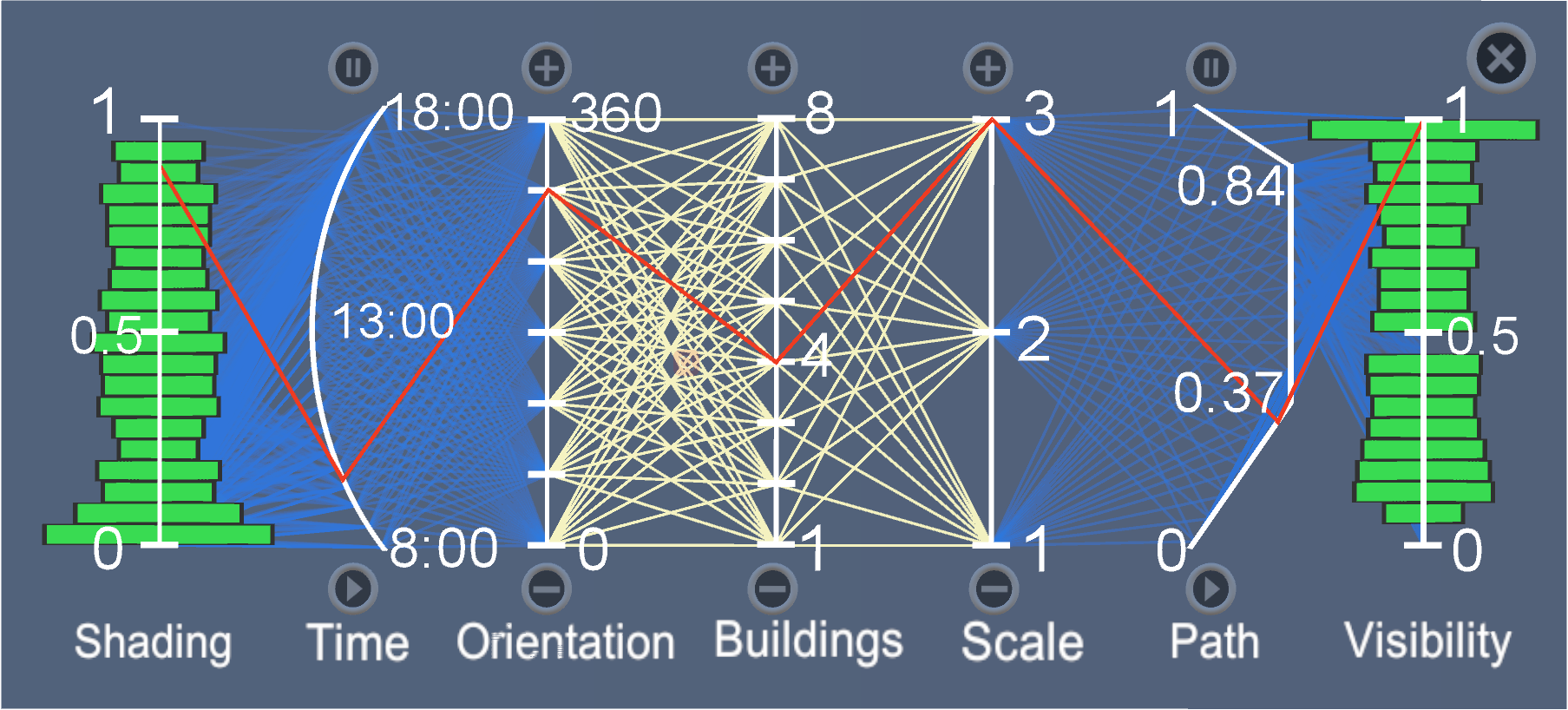}
\caption{Analytics view consists of a customized PCP presenting shadow and visibility analysis results, and various buttons for user interactions.}
\label{fig:view}
\end{figure}
\subsection{Visualization Design}
\label{ssec:vis_design}
The visualization component is designed to support the multi-perspective visualization task (T.2). The component mainly consists of two closely linked views:

\paragraph*{Physical View}
The view presents 3D visualization of an urban area in VR to enable urban context awareness (T.2.1). We employ Unity3D’s built-in VR rendering framework to generate the view. All rendering features are enabled and adjusted to optimal settings, and our system is able to run at interactive frame rates. Selection, manipulation, and navigation operations are enabled in the view through a gesture interaction system described below. Besides, the view also supports animation, which is integrated to facilitate users' understanding of shadow and visibility. For shadow analysis in animation mode, the user’s position will be moved to the optimal viewpoint for the building candidate, and the system’s lighting direction will change over time according to the sun movement. For visibility analysis, the user’s position moves along the architect’s defined path while keeping the target building in view.

\paragraph*{Analytics View}
The view supports visualization of quantitative analysis results (T.2.2), which is mainly made up of a parallel coordinate plot (PCP) as shown in Fig.~\ref{fig:view}. The PCP consists of seven axes, where the middle three correspond to buildings, orientations, and scales of building candidates, the left two correspond to time and shading values for shadow analysis, and the right two correspond to path and visibility results for visibility analysis, respectively. We bend the time axis into an arc to hint sunrise and sundown over time, and arrange the path axis in according to street topology of the path~\cite{8017655}. Since architects would like to first compare building candidates, we integrate bar charts on the left- and right-most axes to indicate shading and visibility distributions for all orientations and scales of a specific candidate. If the candidate’s orientation and scale are further specified, a red line is added in the PCP to show shading and visibility values at certain time and path position.

Besides PCP, various buttons are integrated in the Analytics View. For the middle three axes, plus and minus buttons are placed at the top and bottom, respectively. Users can click the buttons to select a different design, \ie, to change building candidate, orientation, or scale. For time and path axes, start and stop buttons are placed at the top and bottom to control animation. The view is always placed on top of the development site, and if a design is selected, the view will be moved up to the top of the candidate building.

\begin{figure*}[t]
\centering
\includegraphics[width=0.95\textwidth]{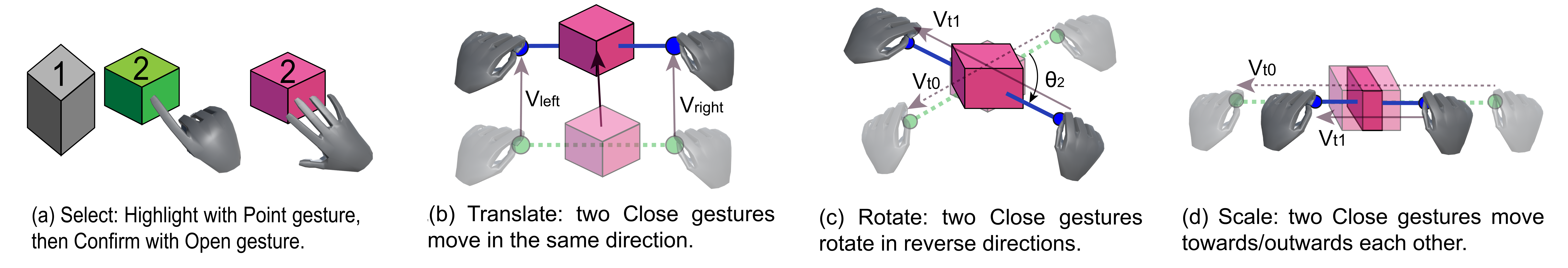}
\caption{Gesture interaction design for interacting with a virtual object: (a) Select, (b) Translate, (c) Rotate, and (d) Scale.}
\label{fig:ges}
\end{figure*}

\subsection{Interaction Design}
\label{ssec:inter_design}

The specific interaction requirements raised by the architect bring in research challenges for interaction design. To support efficient exploration (T.3), we develop a series of egocentric interactions, including robust gesture interaction system, and intuitive handle bar metaphor.

\subsubsection{Gesture Interaction System}
\label{sssec:gis}
To cater to the requirement of efficient and natural interactions, we decided to implement an interaction system based on handle gesture.
We start with a careful study of required operations and raw input gestures.

\paragraph*{Required Operations}
We summarize required operations into the following categories.
\begin{itemize}
	\item \emph{Selection.} Users can select an object in the virtual environment, e.g., to select a development site, or to select a building candidate.
	\item \emph{Manipulation.} After selecting a building candidate, users can manipulate it, including translate its position, rotate its orientation, and scale its size.
	\item \emph{Navigation.} As requested by the architect, our system should allow users to explore the urban area in the virtual space while keeping their body stationary in the real world.
	We opt to seated user postures with gestural interactions for users to navigate the urban scene, which allow users to sit comfortably and suit for most people~\cite{zielasko_2016_evaluation, Zielasko_2020_sitting}. We define two types of navigation operations in our system: pan in x-, y- and z-dimensions, and rotate around y-axis (Unity3D uses y-up coordinate system).
\end{itemize}

\paragraph*{Raw Input Gestures}
Our system detects hand gestures using a Leap Motion device, which is attached to the front of the HMD. When a hand is detected, the device captures various motion tracking data about hands and fingers, including palm position and orientation, fingertip positions and directions, etc. Before deciding what gestures to be used, we first tested different kinds of gestures, including index finger and thumb up, five fingers open, and fist, etc. These gestures are tested with both right and left hands, from different distances, and in different orientations. We identify three most recognizable and stable gestures, i.e., index finger up, five fingers open, and fist. We denote them as \emph{Point}, \emph{Open}, and \emph{Close}, respectively. 

\paragraph*{Gesture Interaction Design}
When one of the three gestures is recognized at time $t$, our system will model the gesture $G_{hand}(t):=<$\emph{Status}, \textbf{P}, \textbf{O}$>$, where $hand \in \{l, r\}$ denoting left and right hand respectively, and \emph{Status} $\in \{Point, Open, Close\}$. \textbf{P} refers to gesture position in the virtual space, which is set to the index finger’s tip position for \emph{Point} and the palm position for \emph{Open} and \emph{Close}. \textbf{O} refers to orientation of the gesture, which is taken as the palm orientation. Orientation is a necessary attribute in our system, as we use orientation stability to classify consecutive gestures belonging to the same operation. Both \textbf{P} and \textbf{O} are 3D vector type data.

We map the raw gestures to interactions as follows.
\begin{itemize}
	\item \emph{Select.} Humans naturally select an object by putting up index finger and pointing at the object. We modify this approach to a two-step operation in our system. First, when a \emph{Point} gesture is detected, our system measures distances from all interactive objects (including building candidates, development sites, and buttons) to the gesture’s position. The closest object with distance less than a threshold will be highlighted. Second, users confirm the selection by opening up the palm, as shown in Fig.~\ref{fig:ges}(a).

	\item \emph{Manipulate.} After an object is selected, users can manipulate it with \emph{translate}, \emph{rotate}, and \emph{scale} operations. These operations also work in two steps. First, left- and right-hand \emph{Close} gestures need to be positioned besides the object at time $t_0$, and our system records the gestures as $G_l(t_0)$ and $G_r(t_0)$, respectively. Second, the gestures will move to other positions at time $t_1$, recorded as $G_l(t_1)$ and $G_r(t_1)$. Corresponding \textbf{P} of these gestures are used to generate two vectors in 3D space:
\begin{equation}
\begin{aligned}
&\textbf{V}_l = \textbf{P}_l(t_1) - \textbf{P}_r(t_0)\\
&\textbf{V}_r = \textbf{P}_r(t_1) - \textbf{P}_r(t_0)
\end{aligned}
\end{equation}
representing movements of left and right hands. 

\emph{Translate:} Next, we measure the angle $\theta_1$ between $\textbf{V}_{l}$ and $\textbf{V}_{r}$. If $\theta_1$ is less than $\pi / 6$, we consider the interaction as \emph{Translate}. A translation corresponding to $\textbf{V}_{trans} = (\textbf{V}_{l} +\textbf{V}_{r}) / 2$ will be applied to position of the object. If the condition is not met, we measure two more vectors:
\begin{equation}
\begin{aligned}
&\textbf{V}(t_0) = \textbf{P}_l(t_0) - \textbf{P}_r(t_0)\\
&\textbf{V}(t_1) = \textbf{P}_l(t_1) - \textbf{P}_r(t_1)
\end{aligned}
\end{equation}

\emph{Rotate:} If $\theta_2$ is above $\pi / 12$, we consider the interaction as \emph{Rotate}.
Buildings can only be rotated around the y-axis. Hence we measure the angle of $\theta_2$ mapped on XZ plane and rotate the building accordingly.

\emph{Scale:} If $\theta_2$ is less than $\pi / 12$, we consider the interaction as \emph{Scale}. The operation is applicable to an object in x-, y-, and z-dimensions. Scaling factor is proportional to $\|\textbf{V}(t_1)\| / \|\textbf{V}(t_0)\|$, and divided in x-, y-, and z-dimensions.
\revise{Notice that scaling is only applicable to a selected virtual object, but not to the whole scene. This is because we would like to maintain the relative proportion between the virtual body and the environment.}

\item \emph{Navigate.} UrbanVR system supports map navigation by camera panning and tilting. The operations are implemented similarly to \emph{Translate} and \emph{Rotate} manipulation operations. The differences include: first, \emph{Open} gesture is used instead of \emph{Close} gesture; and second, no object needs to be selected first.
\end{itemize}

\begin{figure}[!t]
\centering
\includegraphics[width=0.495\textwidth]{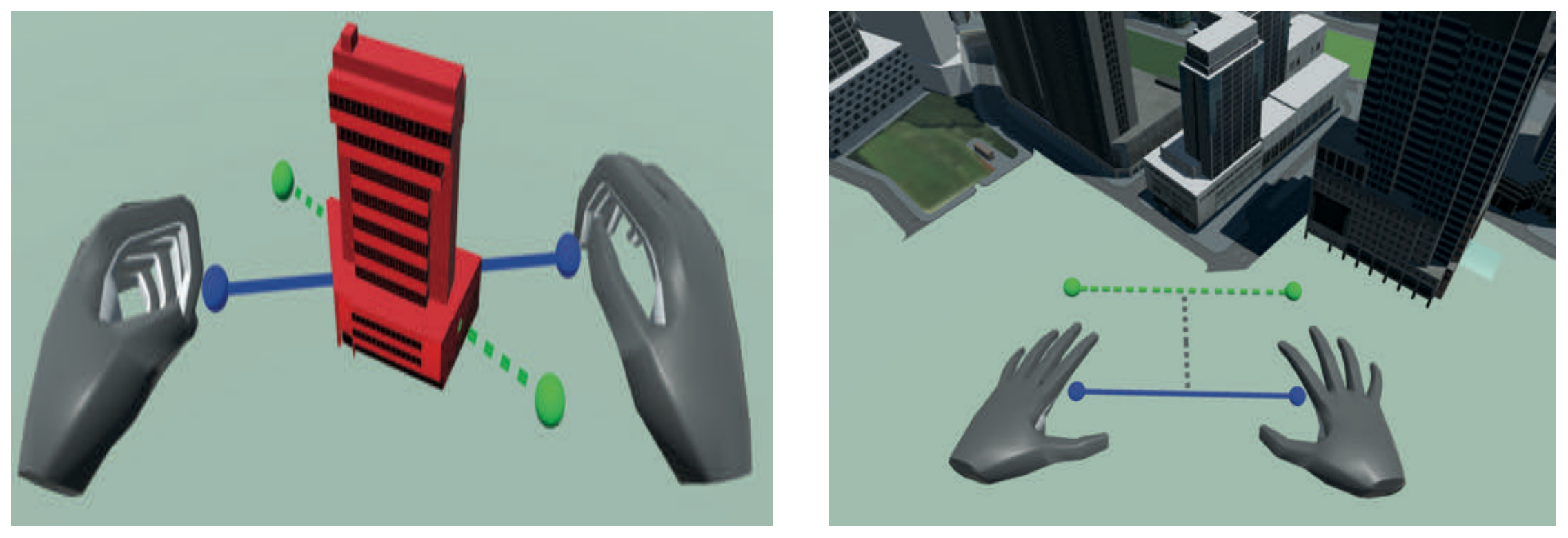}
\caption{Handle bar metaphors representing building rotation (left) and map navigation (right).}
\label{fig:hand}
\end{figure}

\subsubsection{Handle Bar Metaphor}
Continuous visual representations of user operations are necessary for designing effective interactions. Besides choosing the three most accurate gestures, we further improve user interactions through a virtual handle bar.

Handle bar metaphor was proven effective for manipulating virtual objects with mid-air gesture interactions~\cite{song_2012_handle}. Our system adapts this approach: when users are manipulating building candidates or navigating the map, a green and blue handle bar are presented for the initial and moved gesture positions, respectively. First, our system detects if two-hand \emph{Open} or \emph{Close} gestures can be detected for 100 consecutive frames (about one second). Once the first step succeeds, a green handle bar with two balls at both ends and a linking dashed line will be drawn to represent the initial gesture positions. Next, the system detects if follow-up gestures can make up a manipulation or navigation operation, as described in the above section. If an operation is matched, a blue handle bar with a solid connecting line will be presented at the moved gesture positions. Specifically, for \emph{Translate} operations, a gray dashed line connecting middle points of the two handle bars is drawn as well. Fig.~\ref{fig:hand} presents examples of handle bar metaphors representing building rotation (left), and map navigation (right).

\subsection{Viewpoint Optimization}
\label{ssec:viewport}
The complex urban scene can easily cause occlusion that hinders users’ exploration.
Making occluders semi-transparent is not a good solution, since the buildings are coupled with high-resolution textures and translucency of the textures will defect users' perception on the PCP.
Alternatively, users can use gesture interactions described above to navigate the map.
However, map navigation is relatively slow and usually takes a long time to find a good viewpoint. 
This motivates us to develop a method which can help users automatically find an optimal viewpoint. 

Specifically, this work optimizes the camera position that generates maximum view goodness for a target.
The maximum view goodness is defined on the following requirements:

\begin{itemize}
	\item \textit{R1}: Distance and viewing angle to the target shall be not too close that fails to view the whole target, meanwhile not too far away that loses details of the target.
	\item \textit{R2}: Occlusion of the target shall be minimized. Specifically, the camera shall not fall inside any building (R2.1), and there shall be minimum occluders in-between the camera and the target (R2.2).
\end{itemize}
Next, we formulate these requirements as an energy function that can be feasibly solved with optimization algorithms.

\paragraph*{Problem Definition}
In our setup, viewpoint optimization is applied to a target, e.g., the development site or building candidate. We adapt a LoD simplification approach that first simplifies LoD3 buildings in an urban scene into LoD1 cuboids~\cite{Verdie:206788}, and then maximizes the visibility for the target. The target can be represented as $\{CT_i\}_{i=1}^n$, where $CT_i$ represents a cuboid belonging to the target. Target can be occluded by other buildings, and we represent these occluding buildings $\{CO_i\}_{i=1}^m$. Since field of view (FOV) is fixed by the HMD, we can only change the viewpoint position and view direction. We further simplify the problem by always pointing the view ray to the center of the target. Thus, we define the viewport optimization problem as:  
\begin{quote}
{\em 
given cuboids of a target $\{CT_i\}_{i=1}^n$, and cuboids of occluding buildings $\{CO_i\}_{i=1}^m$, to find an optimal position \textbf{P} for observation that generates maximum view goodness of the target.
\/}
\end{quote}

\begin{figure}[t]
\centering
\includegraphics[width=0.495\textwidth]{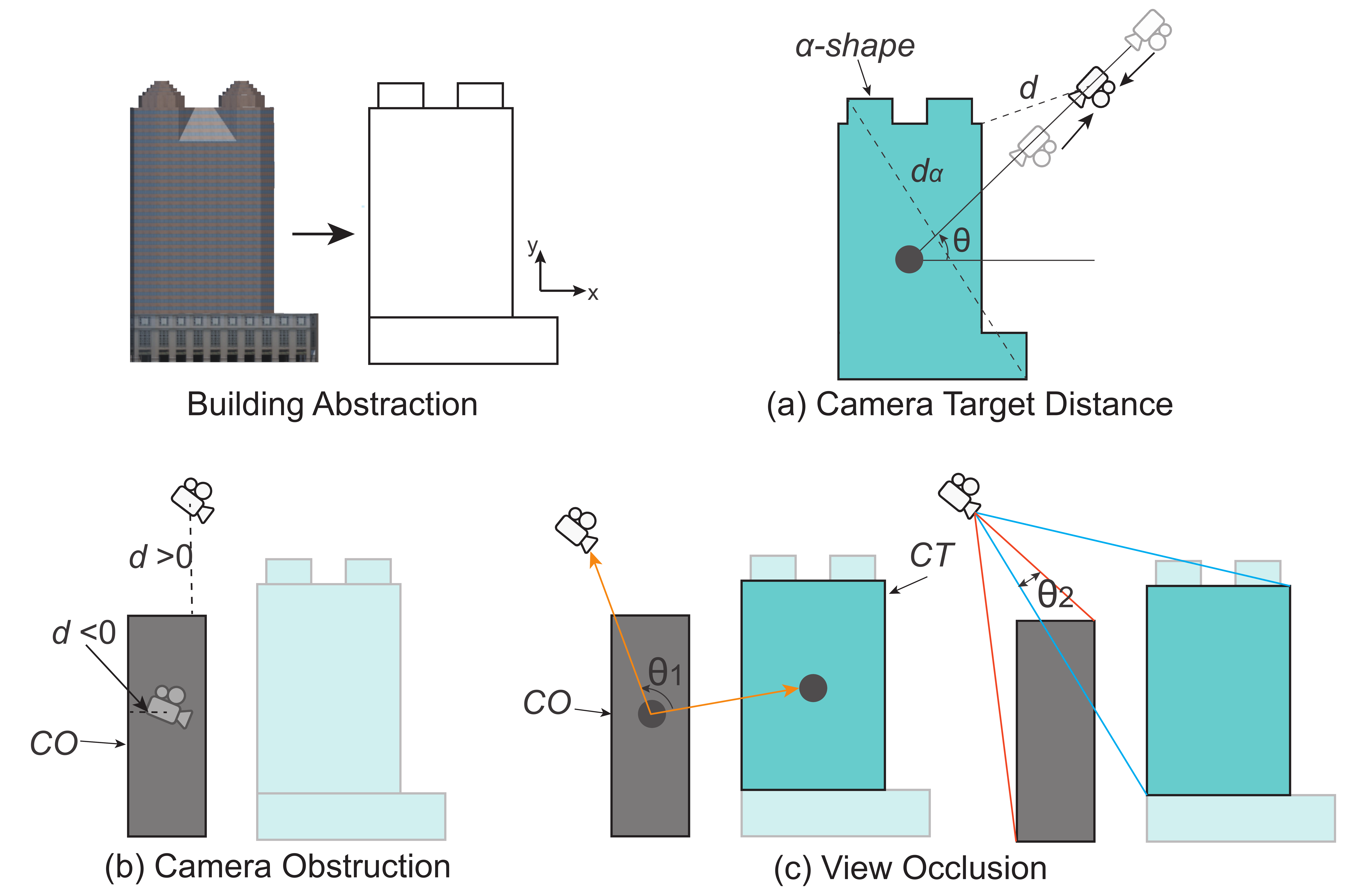}
\caption{Illustration of building abstraction projected to XY plane, and the different energy terms that are used for viewpoint optimization.}
\label{fig:opt}
\end{figure}

\paragraph*{Energy Function}
To solve the problem, we introduce a constraint to optimize \textbf{P} to the target distance (\textit{R1}), a constraint to avoid \textbf{P} inside occluding buildings (\textit{R2.1}), and a constraint to minimize view occlusion of the target (\textit{R2.2}). We formulate these constraints into different energy terms and assemble the energy terms into an energy function. 

Fig.~\ref{fig:opt} presents an illustration of these energy constraints of one building projected onto the XY plane. 
We further repeat projections and measurements on XZ and YZ planes, and combine results from all three planes together. Detailed procedures are described as follows.

\begin{itemize}
	\item \emph{Camera Target Distance ($E_1$).} After projection, we can get a list of vertices in the XY plane from $\{CT_i\}_{i=1}^n$. We extract a closed polygon, i.e., $\alpha$-shape from these vertices, which is the boundary of the target projected onto the XY plane. Maximum diagonal distance of the $\alpha$-shape is calculated and denoted as $d_\alpha$. Then we use a point to polygon distance function which measures distance $d$ from projected viewpoint $\textbf{P}_{xy}$ to the $\alpha$-shape. The distance is negative if the point falls in the closed polygon. To allow arbitrary viewing angles, we introduce $\theta$ which measures the elevation angle of the viewpoint over the horizon from the center of the $\alpha$-shape to the viewpoint. Thus, we create the sub-energy term on the XY plane:

\begin{equation}
E_1(\textbf{P}_{xy}) = \lambda_0 e^{ (d-d_{min}) \times (d-d_{max})} + \lambda_1 e^{ (\theta - \theta_0)^2 + (\theta - \theta_1)^2 }
\end{equation}

where $\lambda_0$ and $\lambda_1$ are the weights for each term. $d_{min}$ and $d_{max}$ are preferred distance range and set to $0.5 \times d_{\alpha}$ and $1.5 \times d_{\alpha}$, respectively. $\theta_0$ and $\theta_1$ are preferred view directions meeting the condition $\theta_0 + \theta_1 = \pi$, and they are empirically set to $\pi/4$ and $3\pi/4$. $E_1(\textbf{P})$ is the sum of all sub-energy terms on each plane. 

	\item \emph{Camera Obstruction ($E_2$).} After projecting $\{CO_i\}_{i = 1}^m$ onto the XY plane, we can get $m$ polygons $\{CO_{xy}^i \}_{i=1}^m$. For each $CO_{xy}^i$, we measure its distance to $\textbf{P}_{xy}$, denoted as $d_i$. The sub-energy term can be constructed as:

\begin{equation}
E_2(\textbf{P}_{xy}, CO_{xy}^i) = e^{d_i}
\end{equation}

To make sure $E_2$ becomes large when \textbf{P} falls in any $CO_i$, we model $E_2$ as:

\begin{equation}
E_2(\textbf{P}) = \sum_{i=1}^{m}{\frac{1}{ \sum{E_2(\textbf{P}_\Pi, CO_\Pi^i)} }}, \Pi \in \{xy, xz, yz\}
\end{equation}

	\item \emph{View Occlusion ($E_3$).} $\{CT_i\}_{i=1}^n$ are projected on the XY plane, and $n$ corresponding polygons $\{CT_{xy}^i\}_{i=1}^n$  are generated. For each $CT_{xy}^i$, we measure how much another $C_{xy}^j$ can occlude it. Here, $C_{xy}^j \in \{CO_{xy}^i \}_{i=1}^m \cup \{ CT_{xy}^i \}_{i=1}^n-\{CT_{xy}^i\}$. The measurement is calculated as follows:

\begin{enumerate}
	\item Generate a vector $\textbf{Q}_{xy}^j\textbf{Q}_{xy}^i$ from center of $C_{xy}^j$ (denoted as $\textbf{Q}_{xy}^j$) to center of $CT_{xy}^i$ (denoted as $\textbf{Q}_{xy}^i$), and a second vector $\textbf{Q}_{xy}^j\textbf{P}_{xy}$ from $\textbf{Q}_{xy}^j$ to $\textbf{P}_{xy}$.

	\item Measure angle $\theta_1$ between $\textbf{Q}_{xy}^j\textbf{Q}_{xy}^i$ and $\textbf{Q}_{xy}^j\textbf{P}_{xy}$. Lager $\theta_1$ means more likely $C_{xy}^j$ may lay in-between $\textbf{P}_{xy}$ and $CT_{xy}^i$.

	\item Generate vectors from $\textbf{P}_{xy}$ to all vertices in $CT_{xy}^i$, and extract two vectors with maximum and minimum angles, denoted as $\textbf{P}_{xy}\textbf{Q}_{xy}^{i0}$ and $\textbf{P}_{xy}\textbf{Q}_{xy}^{i1}$. These vectors represent up- and low-bound view angles for $CT_{xy}^i$, respectively. In the same way, extract $\textbf{P}_{xy}\textbf{Q}_{xy}^{j0}$ and $\textbf{P}_{xy}\textbf{Q}_{xy}^{j1}$ for $C_{xy}^j$.

	\item Lastly, we measure intersection angle $\theta_2$ between $[\textbf{P}_{xy}\textbf{Q}_{xy}^{i0}, \textbf{P}_{xy}\textbf{Q}_{xy}^{i1}]$ and $[\textbf{P}_{xy}\textbf{Q}_{xy}^{j0}, \textbf{P}_{xy}\textbf{Q}_{xy}^{j1}]$. Larger $\theta_2$ means more occlusions.
\end{enumerate}

We construct the energy term as:

\begin{equation}
E_3(\textbf{P}_{xy}) = \sum_{i=1}^{n}{ \sum_{j=1}^{m+n}{\frac{1}{e^{\theta_1} + 1} \times e^{\theta_2}} }, \quad where j \neq i
\end{equation}

$E_3(\textbf{P})$ is the sum of all sub-energy terms on each plane.

\end{itemize}

With these terms defined, we can model the problem as minimization of the following energy:
\begin{equation}
E(\textbf{P}) = \sum_{i=1}^{3}{\omega_i E_i(\textbf{P})}	\label{equ:alg}
\end{equation}
where $\omega_i$ represents the weight for each term, which are empirically set as $\omega_1 =1$, $\omega_2 = 100$, $\omega_3 = 10$, respectively.
$\omega_2$ is the largest to prevent the camera from moving inside a physical object, and $\omega_3$ is larger than $\omega_1$ to minimize view occlusion meanwhile prevent the camera from moving too far away. 

\begin{figure}[t]
\centering
\includegraphics[width=0.495\textwidth]{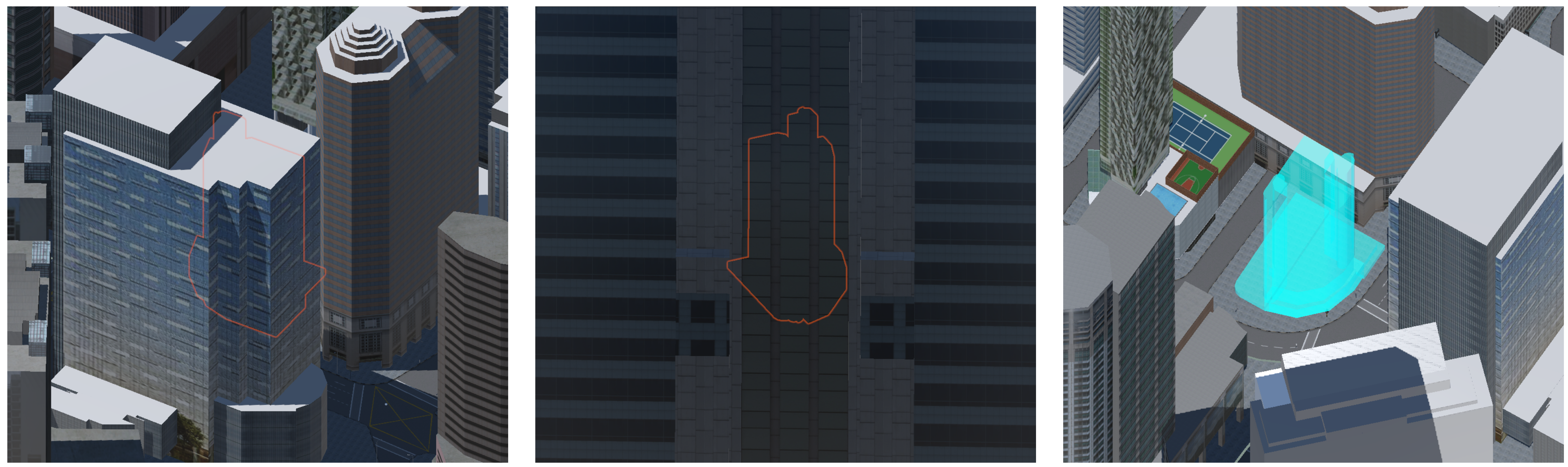}
\vspace{-6mm}
\caption{Effects of different energy terms on viewport optimization. Left: camera target distance (E1) and camera obstruction (E2) only cause occlusion of the target building. Middle: camera target distance (E1) and view occlusion (E3) only block the camera. Right: E1 + E2 + E3 produces optimal viewport for the target building. }
\label{fig:viewOpt}
\end{figure}

Fig.~\ref{fig:viewOpt} illustrates the effects of different energy terms on viewport optimization.
On the left, we adopt the energy function based on defaults weights for camera target distance (E1) and camera obstruction (E2), whilst the weight for view occlusion (E3) is set to zero. The target building (red outline) is obscured by a surrounding building. 
In the middle, we adopt the energy function based on default weights for E1 and E3, whilst the weight for E2 is set to zero. The camera is now positioned inside a building that blocks the target building. 
The full energy function based on E1 + E2 + E3 addresses these issues and leads to optimal viewport as shown on the right.
Note that we neglect the result of energy function based on E2 + E3, since the camera will be located in a far distance.

\subsection{System Implementation}
UrbanVR is implemented in Unity3D. The input models contain 3D geometry information, including geo-positions in WGS-84 coordinate system~\cite{united1987department}, and a third dimension for height. All building models have high-resolution textures, making it suitable for immersive visualization. This, however, also increases computation costs when users interact with our system. In order to accelerate the computation process, we pre-process all building models in LOD3 by abstracting each model into up to five cuboids in LOD1~\cite{Verdie:206788}. The cuboids act as bounding boxes in 3D space for a model. When the system starts, each model is loaded with corresponding cuboids, which are used in viewpoint optimization and gesture interactions.

The viewpoint optimization energy (Equation~\ref{equ:alg}) is a continuous derivative function, which works well for many optimization algorithms. This work employs a Quasi-Newton method. In each computation iteration, the method will find the gradient and length to next position. The process will stop if either a local minimum is found, or the number of iterations is exceeded (1000 in our case). To find multiple local minima, we start 10 parallel computation processes from 10 different initial positions, making full use of the computing resource.
The initial positions are uniformly sampled on the circle that is at a distance of $d_{\alpha}$ (defined in energy term $E_1$) from the center of the target building, and forms a 60 degree angle with the ground above the map.
We also accelerate the optimization process by considering only the buildings near the target building. Specifically, we exclude buildings whose distances are more than $7 \times d_{\alpha}$ from the target building.
After all processes are finished, we combine all candidates and choose the best one.

\section{Evaluation} \label{sec:eva}
UrbanVR is evaluated from two perspectives: First, a quantitative user study is conducted to assess the performance of egocentric interactions and viewport optimization, in accomplishing the task of manipulating a selected building to match the target building. Second, qualitative expert feedbacks associated with a case study in Singapore, are collected for applicability evaluation. 

\subsection{User Study}
\paragraph*{Experiment Design} 
To evaluate the egocentric interactions, we design a within-subjects experiment with 12 conditions: 4 \emph{interaction mode}  $\times$ 3 \emph{scene complexity}. Each participant participated in each condition. For each condition, a target building colored in translucent cyan is positioned in a development site. Participants are asked to select one from multiple building candidates. The selected candidate has different position, orientation, and scale with the target building. Completion time is recorded and used as the evaluation criteria.

\begin{figure*}[!t]
\centering
\includegraphics[width=0.95\textwidth]{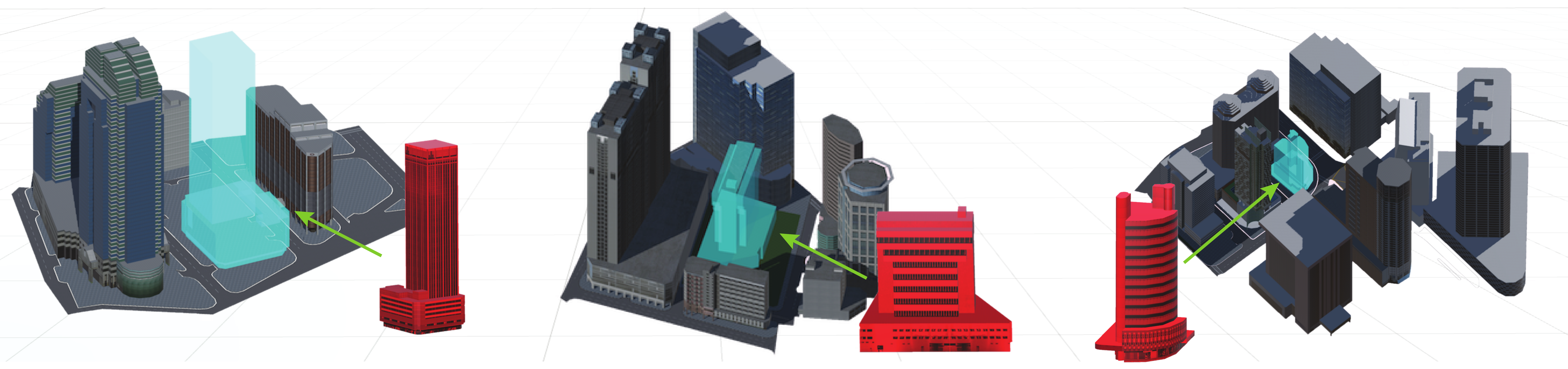}
\caption{Scenes set up in the user study: simple, moderate, and complex from left to right.}
\label{fig:user}
\end{figure*}

\begin{itemize}
	\item \emph{Interaction Mode.} The gesture system provides fundamental interactions for UrbanVR, while viewpoint optimization and handle bar are complementary for improving user experience. We would like to first test if gestures alone work, and then verify if the other two can facilitate user exploration. Hence, we design four modes of interaction:
	\begin{itemize}
		\item C1: Gesture only (Ge).
		\item C2: Gesture and viewpoint optimization (Ge+VO).
		\item C3: Gesture and handle bar (Ge+HB).
		\item C4: Gesture, viewpoint optimization, and handle bar together (Ge+VO+HB).
	\end{itemize}

	\item \emph{Scene Complexity.} In reality, surroundings of a development site can be sparse or dense. High buildings may occlude users’ view of the site, so more interactions are needed to avoid occlusion. This causes more exploration difficulties. As illustrated in Fig.~\ref{fig:user}, we design three scenes with different complexities:

	\begin{itemize}
		\item S1: Simple. The surrounding consists of only two buildings; Fig.~\ref{fig:user} (left).
		\item S2: Moderate. The site is surrounded by six sparsely located buildings. Two of them are relatively low height on one side, from where the space can be viewed without occlusion; Fig.~\ref{fig:user} (middle).
		\item S3: Complex. The surrounding consists of eight densely located buildings. All buildings are higher than the target building, thus the space is occluded from almost all viewpoints; Fig.~\ref{fig:user} (right).
	\end{itemize}
\end{itemize}

\paragraph*{Participants}
We recruited 15 participants, 10 males and five females. 12 of them are graduate students, and the others are research staff. The age of the participants range between 20 and 30 years. Two participants played VR games using the HTC Vive controllers. No participant has experience with gesture interactions in VR before the study. Three participants have a background in architecture, while the others have no knowledge about urban design.

\paragraph*{Apparatus and Implementation}
UrbanVR system was implemented in Unity3D.
The experiments were conducted on a desktop PC with 12 $\times$ Intel(R) Core(TM) i7-6800K CPU @ 3.4GHz, 32GB memory, and a GeForce GTX1080 Ti graphics card.
The VR environment was running in a HTC Vive VR HMD, and the hand gestures were captured using a Leap Motion attached to the front of HMD.

\paragraph*{Procedure}
The studies are performed in the order of introduction, training, experiment, and questionnaire. First, we present a 5-min introduction about the interactions, followed by a 10-min training session to make sure all participants are familiar with the interactions. Then, the experiment starts, and completion time for each experiment condition is recorded. In the end, feedbacks are collected. The questions include if they had experience with gesture interactions in VR, if they feel dizzy, if they think the gestures are natural, and advices for improvement.

In each experiment condition, the starting viewpoint is positioned on top of the urban scene. Participants are asked to complete the task requiring the following operations:

\begin{itemize}
	\item \emph{Navigate to the site.} This can be either done through gesture-based map navigation, or viewpoint optimization by selecting the site.
	\item \emph{Select a building.} Participants can open up a menu through left-hand Open gesture, and eight building candidates will be presented at the left hand position. Then participants can select a building that matches the target through right hand Selection. The selected building will be placed besides the menu.
	\item \emph{Manipulate the selected building to match the target.} Position, orientation, and size of the selected building differ from the target. Participants need to manipulate the selected building to match the target.
\end{itemize}

All operations are completed while the participants are sitting in a chair. This process is repeated in total 12 times (4 interaction combinations $\times$ 3 scene complexities) for each participant. The participants are asked to take a break (3 minutes) in every 10 minutes, and take a break (5 minutes) after finishing a task. Sequence of the interaction combination is pseudo-randomly assigned to each participant in order to suppress learning effects gained from previous assignments.
Each participant was compensated with SGD \$40.00 ($\sim$USD \$30) after the study.

\paragraph*{Hypotheses}
We postulate the following hypotheses:

\begin{itemize}
	\item H1. Gesture only (C1) is the slowest interaction technique.
	\item H2. Viewpoint optimization (C2) can facilitate gesture only (C1) interaction, \ie, time(C1) $>$ time(C2).
	\item H3. Handle bar metaphor (C3) can facilitate gesture only (C1) interaction, \ie, time(C1) $>$ time(C3).
	\item H4. All techniques together (C4) is the fastest technique, \ie, time(C2) $>$ time(C4) and time(C3) $>$ time(C4).
\end{itemize}

\paragraph*{Experiment Results}
15 experiment results are generated for each experiment condition. We first test the results in each condition against normal distribution using Shapiro-Wilk test. The results show that conditions of C2 do not follow normal distribution, while the results in other conditions are following normal distributions under certain probability.

Prerequisites for computing ANOVA are fulfilled for H1, H3 and H4. We perform a two-way ANOVA (3 interaction combinations $\times$ 3 scene complexities) on them. Significant effects of interactions on completion time $(F_{(2,126)}=64.97, p < 0.001, \eta^2 = 1.718)$ are observed. Scene complexity also has a significant main effect on completion time $(F_{(2,126)}=59.63, p < 0.001, \eta^2 = 1.646)$. We use Kruskal-Wallis test~\cite{ref1} to evaluate H2. The result shows the viewpoint optimization has significant effect on completion time $(H_{(2)}=25.53, p < 0.001, \eta^2 = 0.560)$.

\begin{figure}[!t]
\centering
\includegraphics[width=0.485\textwidth]{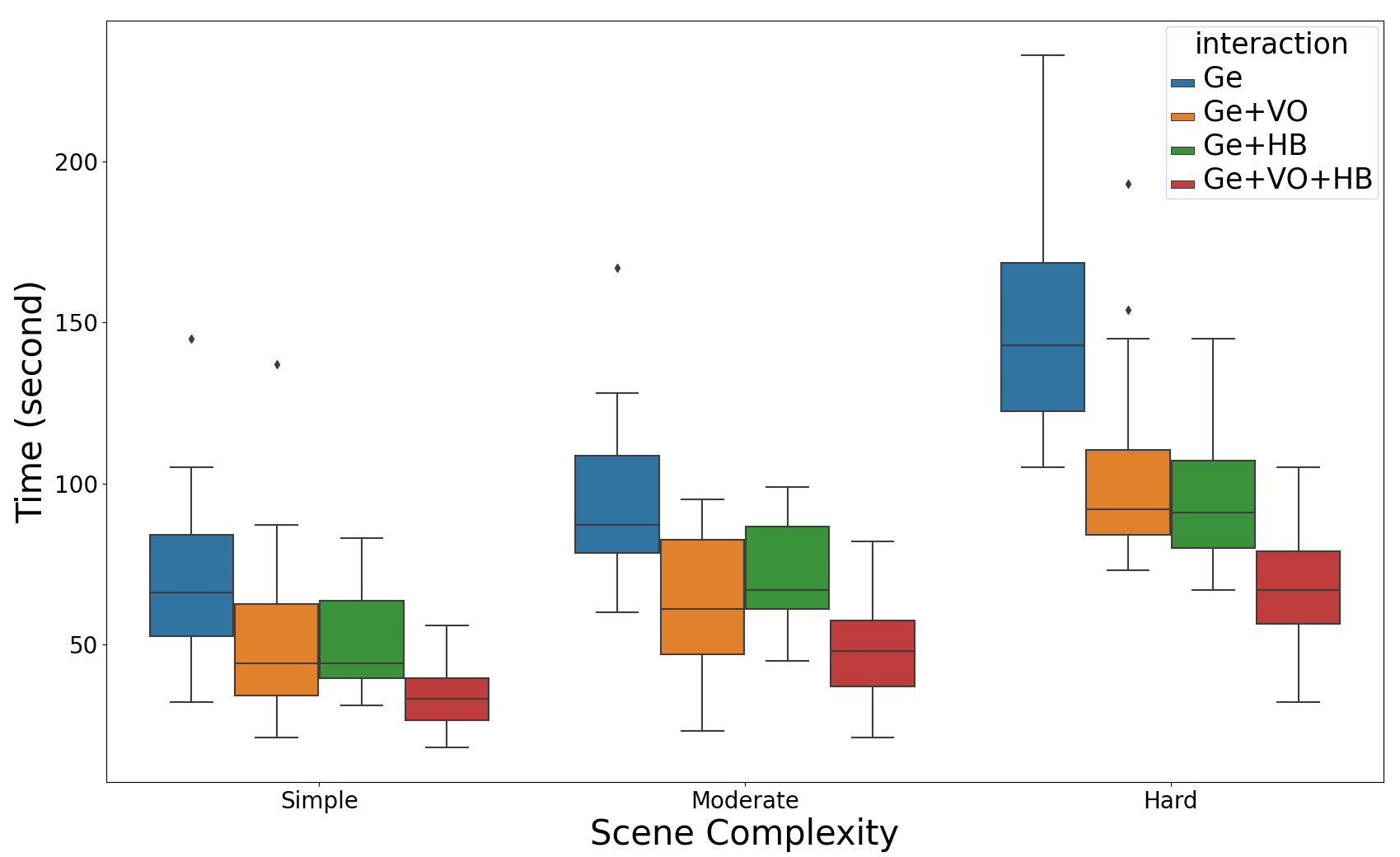}
\caption{Task completion time (second) for each condition.}
\label{fig:task}
\end{figure}

We also perform post hoc comparisons of completion times among the interaction combinations. 
The results are shown in Fig.~\ref{fig:task}. C1 is on average more than 33s slower than C3 $(t_{(44)} =33.42,\ p < 0.001,\ d_{Cohen} = 0.882)$, while C3 is on average 23s slower than C4 $(t_{(44)} =22.84,\ p < 0.001,\ d_{Cohen} = 0.955)$;
C1 is on average more than 32s slower than C2 $(t_{(44)} =32.38,\ p < 0.001,\ d_{Cohen} = 0.772)$, while C2 is on average 24s slower than C4 $(t_{(44)} =23.89,\ p = 0.001,\ d_{Cohen} = 0.799)$.
The results confirmed H1, H2, H3, and H4. Through more detailed probes, we figure out: 
C1 is slower than C3 $(t_{(14)} =21.07 ,\ p = 0.018,\ d_{Cohen} = 0.921),\  (t_{(14)} =23.93 ,\ p = 0.011,\ d_{Cohen} = 1.032),\ (t_{(14)} =55.37 ,\ p < 0.001,\ d_{Cohen} = 1.698)$, 
while C3 is slower than C4 $(t_{(14)} =16.73 ,\ p = 0.003,\ d_{Cohen} = 1.225),\ (t_{(14)} =23.20 ,\ p = 0.002,\ d_{Cohen} = 1.336),\  (t_{(14)} =28.60 ,\ p = 0.002,\ d_{Cohen} = 1.399)$;
C1 is slower than C2 $(t_{(14)} =19.20 ,\ p = 0.119,\ d_{Cohen} = 0.656),\  (t_{(14)} =32.067 ,\ p = 0.035,\ d_{Cohen} = 1.284),\ (t_{(14)} =45.87 ,\ p = 0.156,\ d_{Cohen} = 1.242)$, 
while C2 is slower than C4 $(t_{(14)} =18.60 ,\ p = 0.148,\ d_{Cohen} = 0.816),\ (t_{(14)} =15.07 ,\ p = 0.170,\ d_{Cohen} = 0.765),\  (t_{(14)} =38.00 ,\ p < 0.031,\ d_{Cohen} = 1.413)$ for S1, S2, and S3, respectively. 
We use False Discovery Rate~\cite{https://doi.org/10.1111/j.2517-6161.1995.tb02031.x} ($\alpha = 0.05$) for the correction of above data. The results suggest that for more complicated scenes, handle bar metaphor and viewpoint optimization techniques make interactive VR exploration more efficient.

\begin{figure*}[!t]
\centering
\includegraphics[width=0.85\textwidth]{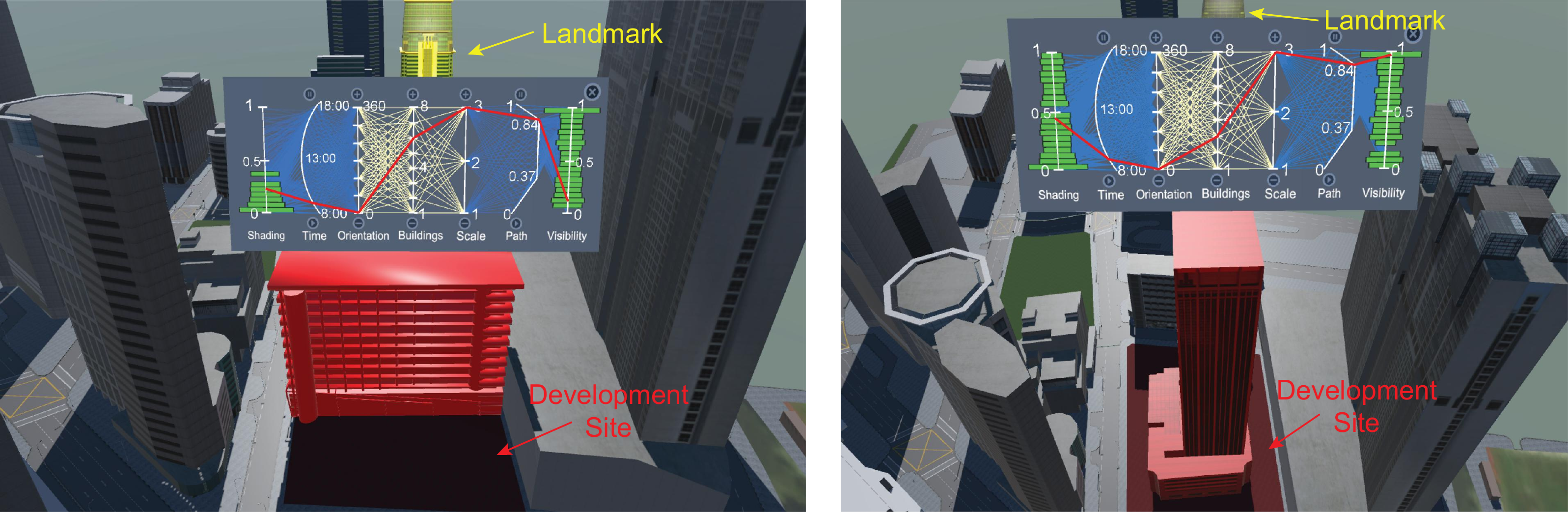}
\caption{Two building candidates are evaluated against shading and visibility using UrbanVR.}
\label{fig:eva}
\end{figure*}

\paragraph*{User Feedback}
All participants finished the experiment in 1.5 hours. No one felt dizzy with our system in 10-min tests, but one felt a bit eye dry. All participants agreed that the gestures are natural and easy to use, and viewpoint optimization and handle bar are quite helpful. Three participants especially liked the viewpoint optimization, as they got easily bored and tired when navigating the map using gestures. ``\emph{Viewpoint optimization is really cool. I think too much navigation just makes users feel bored.}" There are also some negative feedbacks. Two participants felt the HMD resolution is not high enough, which reduces immersive feeling. One participant was not able to manipulate the building precisely, even with the handle bar. It took her a long time to complete the task.

\subsection{Expert Feedbacks}
To evaluate the applicability of UrbanVR in urban design, we further conducted expert interviews with two independent architects (denoted as EA $\&$ EB) other than our collaborating architect. EA is a registered architect in Germany with more than 15 years of experience, while EB has about five years experience.
In the interviews, we started by explaining the tasks, visual interfaces, and interaction designs, and demonstrated a case study of how UrbanVR can be utilized in a real-world scenario in Singapore. The following data are received from the collaborating architect.
\begin{itemize}
	\item 3D models of a central business district in Singapore. The district is about 1 km long and approximately 0.5 km wide. The models include about 50 buildings, and a number of other objects, such as roads and street furniture.
	\item Eight building models that are used as building candidates.
\end{itemize}

In the study, we first specify a development site and a landmark building at back for visibility analysis. Visibility and shadow analysis results are precomputed for the building candidates. Next, we select candidates and evaluate their visibility and shading impacts. Two candidates are compared in Fig.~\ref{fig:eva}.

\begin{itemize}
	\item The left figure presents a low and wide candidate. From the bar chart on the shading axis, we can see that the shading values are concentrated in between [0, 0.5], indicating the candidate generates little shadow on the surrounding. From the bar chart on the visibility axis, the values are approximately evenly distributed in between [0, 1], except for a high concentration at 1. If we further specify the view position at a street corner (the turning point on the path axis), we can see the visibility is low, which indicates that the candidate occlude the landmark a lot at the corner.
	\item A tall and thin candidate is shown in the right figure. In comparison with Fig.~\ref{fig:eva} (left), the bar distribution on the shading axis shows relatively more values over 0.5, and visibility values are more concentrated at 1. This indicates that the candidate generates more shading while affect less visibility than the left one. In addition, the red line also indicates that the candidate does not occlude the landmark at the corner.
\end{itemize}

After the demonstration, the experts were asked to explore the system by themselves for about 20 minutes. Both of them repeated the case study scenario successfully. Feedbacks about the system were collected and summarized below.

\paragraph*{Feasibility}
The experts agreed that VR is a new technology that is certainly worth exploration for urban design. Both experts appreciated our work in applying immersive analytics for urban design. They know some planning teams are exploring VR and AR technologies. However, these works are more ``\emph{focusing on demonstrations and lack analysis}", according to EA. In comparison, our system well integrates visualization and analysis. EB also commented that UrbanVR provides ``\emph{immersive feelings of an urban design, and more importantly, being able to modify the plan and provide quantitative results}".

\paragraph*{Visual Design}
Both experts had no motion sickness with the visualization. They are familiar with the urban scene studied in the work, yet they did not expect to get much more immersive feelings in VR than desktop 3D visualizations. ``\emph{The animation is very vivid, especially when navigating on the path}" in the visibility analysis, commented by EB. The experts had some difficulty in understanding the PCP in the beginning. EA thought some simple line charts would be more effective. He was persuaded after we explained that there could be too much line charts to represent all visibility and shadow analysis. They liked the idea of bending time-axis into a curve, and arranging path-axis in the same topology with the streets. ``\emph{These small adjustments give me strong visual cues of reality}", commented by EB. The study clearly reveals that for the development site, shading is determined by building heights, while visibility is mainly affected by building width. The collaborating architect appreciated this finding and praised the bar chart design on the shading and visibility axes.

\paragraph*{Interaction Design}
Both experts got used to the gesture interactions quickly. They tried both building manipulation and map navigation using gestures, and felt the gestures are easy to use in VR. EA especially liked the handle bar metaphor, which ``\emph{helps a lot when manipulating the buildings}". In the beginning, the experts did not realize that when visualizing the case study, the viewpoint was optimized. After the demonstration, they liked the idea and felt it is necessary, as EA felt ``\emph{the HMD is too heavy - not suitable for long-time wearing}".

\paragraph*{Limitations}
The experts pointed out two major limitations in our system. First, the experts felt the ``\emph{gestures are not comparable with mouse regarding accuracy}", even though we have provided handle bars for visual feedbacks. Nevertheless, they liked gestural interactions because hand guestures are natural, and they encouraged us to further improve the accuracy. Second, the experts would like to see more analysis features integrated into our system, such as building functionality, street accessibility, transportation mobility, etc. The current analyses are not covering all necessary evaluation criteria they need.
\section{Conclusion, Discussion, and Future Work}
We have presented UrbanVR $-$ an immersive analytics system that can facilitate site development. UrbanVR integrates a GPU-based image processing method to support quantitative analysis, a customized PCP design in VR to present analysis results, and immersive visualization of an urban site.
In comparison to similar tools on the desktop (\eg,~\cite{ferreira_2015_urbane, ortner_2016_vis-a-ware, 8283638}), the main advantage of UrbanVR is that the immersive environment gives users the feeling of spatial presence of ``being there".
This is especially appreciated for context-aware urban design, as commented by the experts.
Nevertheless, the immersive environment also brings about difficulty for user interaction.
We develop several egocentric interactions, including gesture interactions, viewpoint optimization, and handle bar metaphors, to facilitate user interactions. The results of the user study show that viewpoint optimization and handle bar metaphor can improve gesture interactions, especially for more complex urban scenes.
There is an emerging trend for immersive data analysis in various application domains. In addition to general immersive visualizations, customized designs that can fulfill specific analytical tasks, are also in high demand.

\paragraph{Discussion}
There is a strong desire for immersive analytics systems that can facilitate urban design using personalized HMDs. Examples can be found in both virtual (\eg~\cite{Liu_2020_three, schrom2020interactive}) and augmented reality (\eg~\cite{noyman2019cityscopear, Ariel_2020_deepscope}) environments. However, challenges exist in every stage of system development, from task characterization to data analysis to visualization and interaction designs. Through the development of UrbanVR, we gain insightful experience. First, a close collaboration with domain experts is necessary from the beginning, and iterative feedbacks from domain experts can help reduce unnecessary efforts. Second, integration of advanced techniques from cross domains of visualization, graphics, and HCI fields can greatly improve usability of the system. 

Currently, limited user interactions are supported by UrbanVR.
\revise{In the gesture interaction design (Sect.~\ref{sssec:gis}), we do not implement scene scaling, but only provide camera panning.
Even though both interactions can make the scene look bigger or smaller, the experience they bring to the user is completely different in the VR environment, and this affects user behavior~\cite{Zhang2009, 7460054}.
}Besides, more advanced interactions for exploring analysis results, such as to filter and reconfigure the PCP~\cite{4376144}, are missing. This limits analysis functionality of our system. For instance, the collaborating architect would like our system to filter and sort designs according to a specific criteria. UrbanVR will be more useful if such features are integrated. A main obstacle is that the raw gesture detection provided by Leap Motion is not accurate enough. We look forward to more advanced interaction paradigms, such as hybrid interactions that combine gestural and controller-based interactions~\cite{besancon_2021_star}, and more accurate interaction algorithms, such as deep learning techniques~\cite{chen_2020_lassonet}, in the near future. With more accurate and robust gesture interactions, we can further improve the interface design, following the guidelines for interface design for virtual reality~\cite{BOWMAN199937}.

For the moment, we integrate viewpoint optimization and handle bar metaphor to improve gesture interactions. The user study shows that viewport optimization (C2), handle bar (C3), and handle bar + viewpoint optimization (C4) facilitate gesture interaction in terms of completion time. The tasks require users to match a building’s position, size, and orientation with a target building. Hence, the matching can also be considered as an accuracy test. In terms of this, the handle bar also improves the accuracy of gesture interactions. Results for viewpoint optimization as a supplement to gesture interactions do not distribute normally. This may be because the participants are not familiar with viewpoint optimization. More try-outs may suppress the participant’s cognitive bias on viewpoint optimization.

Besides viewpoint optimization, another popular occlusion minimization and view enhancement method is $focus + context$, which has also been well-studied in the visualization and graphics communities. Recently, some study has applied this method to urban scenes~\cite{7208899}. The method allows buildings to be shifted and scaled, thus should generate views with less occlusion than our approach. However, the distortion of urban scene may increase the burden of mental mapping of reality in the virtual world. Considering the benefits and drawbacks, it is worth another study to compare the effectiveness of these methods for immersive analytics.

\paragraph{Future Work}
There are several directions for future work. First, we would like to implement the $focus + context$ technique in VR, and compare it with the viewpoint optimization method to check which one is more effective. In addition, we will continue collaborating with domain experts, to find more problems and data that are suitable for immersive analytics.
To facilitate applicability of UrbanVR, we plan to add support for common data formats of urban planning, and integrate an export function that allows users to export their designs, in the near future.
Nevertheless, the system is expected to encounter scalability issues when the tasks become complex and data become diverse. This will bring in new research challenges and opportunities for immersive analytics.


\section*{Acknowledgments}
This work is supported partially by National Natural Science Foundation of China (62025207) and Guangdong Basic and Applied Basic Research Foundation (2021A1515011700).

\bibliographystyle{cag-num-names}
\bibliography{realRefs}

\end{document}